\documentclass[aps,prl,a4paper,11pt,amsfonts,amssymb]{article}
\usepackage{amsmath}
\usepackage{amsfonts}
\usepackage{amssymb}
\usepackage{bbm}

\begin{document}

\def\ket#1{|\,#1\,\rangle}
\def\bra#1{\langle\, #1\,|}
\def\braket#1#2{\langle\, #1\,|\,#2\,\rangle}
\def\ketbra#1#2{\ket{#1}\bra{#2}}
\def\identity{\leavevmode\hbox{\small1\kern-3.8pt\normalsize1}}  

\def\lineacrosspage{\hbox to \hsize{\hfill\rule[5pt]{2.5cm}{0.5pt}\hfill}}
\def\FLAG{ \par \medskip \noindent \emph{[This Section Not Yet Complete]} \par \medskip }
\def\comment#1{}

\def\set#1{\{ #1\}}
\def\Prob#1{\mbox{Prob}(#1)}
\def\modulus#1{\left| #1 \right|}
\newcommand{\QED}{\nopagebreak\hspace*{\fill}\mbox{\rule[0pt]{1.5ex}{1.5ex}}}
\newcommand{\qed}{\mbox{\rule[0pt]{1.5ex}{1.5ex}}}
\newcommand{\half}{\mbox{$\textstyle \frac{1}{2}$} }
\def\indicator#1{\left\{ \phantom{\big|} #1 \phantom{\big|}\right\}}

\newtheorem{theorem}{Theorem}
\newtheorem{definition}{Definition}
\newtheorem{lemma}{Lemma}
\newtheorem{example}{Example}
\newtheorem{property}{Property}
\newtheorem{proposition}{Proposition}
\newtheorem{corollary}{Corollary}
\newtheorem{conjecture}{Conjecture}

\title{Computation with Unitaries and One Pure Qubit}
\author{D.~J.~Shepherd\footnote{shepherd@compsci.bristol.ac.uk,  dan.shepherd@cesg.gsi.gov.uk}\\
\small{\it University of Bristol,  Department of Computer Science}\\
}
\date{November 28, 2006}
\maketitle

\vspace{-0.3 cm}
\abstract{
We define a semantic complexity class based on the model of quantum computing with just one pure qubit (as introduced by Knill \& Laflamme) and discuss its computational power in terms of the problem of estimating the trace of a large unitary matrix.  We show that this problem is complete for the complexity class, and derive some further fundamental features of the class.  We conclude with a discussion of some associated open conjectures and new oracle separations between classes.
} \vspace{0.5 cm} \normalsize

\section{Introduction} \label{sect:intro}
The idea of doing quantum computation with just one pure qubit is relevant to certain problems in the theory of quantum chaos~\cite{lit:KL98}, and possibly relevant in terms of choosing an architecture to pioneer practical quantum computation, since at present there would seem to be little agreement about which technology is most fit for implementing a general purpose quantum processor,~\cite{book:NandC}.  Yet the literature seems to lack a discussion of the complexity classes naturally associated to the one-pure-qubit paradigm.  Taking a computer science perspective, we will define a class analogous to {\bf BQP} using the circuit model, discuss some basic conjectures about that class, provide a definition of relativisation for the model, and show some evidence for the conjectures in terms of that relativisation.  It is hoped that this will help promote the search for algorithms in other paradigms.

DQC1 as a computational model, (``Deterministic\footnote{The adjective ``Deterministic'' will not prove to be especially relevant throughout.} Quantum Computation -- 1 pure qubit'', as defined in~\cite{lit:KL98}) is an apparently less powerful computational idea than `full' quantum computation, yet it appears to be able to do certain things exponentially more quickly than can a classical device.  In particular, the original paper~\cite{lit:KL98} explains something of the relation to the classically hard problem of trace estimation.  In this article, we use the idea of computational circuits to derive a proper complexity class for the DQC1 model, in order to ask about the languages that can be resolved with bounded probability in polynomial time, and we address conjectures concerning the ability of families of such circuits to resolve some of the `classical' languages belonging to classes such as {\bf P} and {\bf BPP}.

There is a nice description of general quantum circuits in~\cite{lit:TD0205}, where it is shown how circuits can naturally be composed whenever the output of one circuit is of the same ``type or kind'' as the input to the next.  The authors also point out that it can sometimes be useful to have the elements of a quantum circuit depend on a classical string that has the status of `input', to allow for notions of adaptation.  Both of these themes are key to the development of a computational class that expresses the power of the DQC1 model.

This first section recaps some basic notation used ubiquitously within Quantum Information Science, recaps some definitions of \emph{circuits} as used in context of computational complexity theory, and introduces a new complexity class for capturing the power of the DQC1 methodology within the context in which we're interested.
Section \ref{sect:completeness} illustrates in detail a well-known complete problem for the ``one pure qubit'' approach, and shows how our complexity class is relevant to that problem.
Section \ref{sect:properties} looks in more detail at the new class, proving some basic properties about it and showing that it is closed under certain kinds of reduction. 
Section \ref{sect:conjectures} makes some conjectures about how our complexity class stands in relation to other common classes, analysing the strength of the conjectures with regard to known hard challenges of complexity theory, and providing `evidence' for believing the conjectures despite an absence of formal proof.

\subsection*{Algebra Notation} \label{notation}
We use the algebra formally defined by 
\begin{equation}
  {\cal A}_w ~:=~ \frac{ {\mathbbm C}\bigl[~ X_i, Z_i ~\bigr]_{i=1}^w }
                       { \bigl<~ X_i^2-1, Z_i^2-1, (X_i, Z_i), [X_i, X_j], [X_i, Z_j] ~\bigr>_{i\not=j} },
\end{equation}
where $w$ is a positive integer, the \emph{width} (in qubits) of a quantum circuit.
This algebra is nothing other than the ring of square ($2^w$-by-$2^w$) complex matrices, but it is useful to be able to refer to the algebra without having to speak explicitly of matrices.  On it we can define Trace ($Tr$) and adjoint ($\dag$) in the usual manner, these having the properties that one would expect of a matrix algebra.  (Note that the value of $Tr[1]$ will depend on which ${\cal A}_w$ the $1$ is taken from.)  
$X$ and $Z$ denote the usual Pauli operators.
The unitary transformations of quantum physics correspond with the automorphisms $~\rho \mapsto U \cdot \rho \cdot U^\dag$, for unitary $U$.

The DQC1 model is characterised by requiring the starting state of any computation to be highly mixed except for one pure qubit~: the starting state is given as the (Hermitian) density operator 
\begin{equation}  \label{eqn:startstate}
  \rho_{\mbox{  \footnotesize{\!start}}} ~:=~ \frac{1 + Z_1}{2^w} ~\in~ {\cal A}_w.
\end{equation}

The formalism must expressly forbid any non-unitary gates, since the use of `zeroise' gates, for example, will readily enable one to purify more qubits and then perform a computation on many pure qubits.  Instead, the \emph{only} allowed actions on the density operator (within DQC1,) besides a final measurement, are conjugations by unitary operators (\emph{i.e.} inner automorphisms.)  
Thus the states (density operators) reachable from the starting state using only these allowed automorphisms have the form
\begin{equation}  \label{eqn:genericDQC1state}
  \rho ~=~ \frac{1+UZ_1U^\dag}{2^w} ~=:~ \frac{1 + \beta Z_1 + \sqrt{1-\beta^2}R}{2^w},
\end{equation}
where $R$ and $Z_1 R$ are traceless, and $\beta$ is real.
This format for writing a generic state is chosen so as to highlight the coefficient in front of the $Z_1$ term, which we will later define as the `output' of a circuit whose gates implelemt the unitary $U$.
That $\beta$ is real can be seen from its definition as $~\beta := Tr(U Z_1 U^\dag Z_1)/2^w~$ and the fact that $U Z_1 U^\dag Z_1$ is a product of two hermitian operators.\footnote{$(AB)^\dag = B^\dag A^\dag = BA$; $\quad Tr[BA] = Tr[AB]$.}
That $R$ and $Z_1 R$ are traceless can be seen from the definition $~R := \frac{U Z_1 U^\dag - \beta Z_1}{\sqrt{1-\beta^2}}$.  (We leave $R$ undefined in the cases where $\beta^2=1$.)

We think of $R$ as coding the state of the `workspace' of a DQC1 algorithmic process, and $\beta$ as coding the `output' of the process.
The final measurement, to be applied to the state after the completion of the unitary circuit, will measure the first qubit in the computational basis.  The probability of obtaining ``0'' from this reading for the state described in equation~(\ref{eqn:genericDQC1state}) is therefore given by
\begin{equation} \label{eqn:probability}
  \mathbbm{P}(\mbox{``0''}) ~=~ Tr\left[ \rho \cdot \frac{1+Z_1}2 \right] ~=~ \frac{1+\beta}{2}.
\end{equation}

\subsection*{Entanglement}

It is worth mentioning here something about the entanglement of states of the form of equation~(\ref{eqn:genericDQC1state}).  While we won't be considering any particular applications of entanglement, we note that entropy considerations alone guarantee that, for any reasonable measure, there won't be \emph{much} entanglement present in such a system.  But it would be inaccurate to say that such states are necessarily free of \emph{all} quantum correlations.  An example of an entangled state would be
\begin{equation} \label{eqn:entangledexample}
  \rho_{ \mbox{\footnotesize{Ent}}} ~:=~ \frac1{2^{w+1}}(~ 2 + X_1 X_2 - Y_1 Y_2 + Z_1 - Z_2 ~),
\end{equation}
which is prepared from the state $\rho_{\mbox{  \footnotesize{\!start}}}$ by the unitary operator
\begin{equation} \label{eqn:entangledunitary}
  U ~=~ \frac12(~ 1 + Z_1 + X_2 - Z_1 X_2 ~) \cdot \frac12(~ 1 - Z_2 + H_1 + H_1 Z_2 ~).
\end{equation}
To see that it is entangled, note that $\rho_{ \mbox{\footnotesize{Ent}}}$ satisfies
\begin{eqnarray}
  Tr[~ \rho \cdot (1-Z_1)(1+Z_2) ~] &=& 0,            \label{eqn:firstrule} \\
  Tr[~ \rho \cdot (X_1+iY_1)(X_2+iY_2) ~] &\not=& 0.  \label{eqn:secondrule}  
\end{eqnarray} 
However, any pure product state 
\begin{equation}
  \frac1{2^w}(1+\alpha_1X_1 + \beta_1Y_1 + \gamma_1Z_1) \cdot (1+\alpha_2X_2 + \beta_2Y_2 + \gamma_2Z_2) \cdots
\end{equation}
either satisfies equality~(\ref{eqn:firstrule}) and then fails the inequality~(\ref{eqn:secondrule}), or else satisfies the inequality~(\ref{eqn:secondrule}) but then fails the equality~(\ref{eqn:firstrule}) by contributing a strictly positive (real) amount.  Therefore $\rho_{ \mbox{\footnotesize{Ent}}}$ cannot be a convex combination of pure product states.

\subsection*{Circuits in general}
Here we recap a few basic definitions for circuits and computation via circuits.
A \emph{classical} deterministic circuit is usually defined (\emph{e.g.} \cite{book:Pap}) as a directed acyclic graph whose vertices are either `inputs' or `constants' (with no indegree), `NOT gates' (with indegree 1), or `AND' or `OR' gates (with indegree 2).  The vertices with no outdegree are called `outputs'.  The language obtained from an infinite family $\{ C_1, C_2, ... \}$ of such circuits, where circuit $C_n$ has $n$ inputs and one output vertex, is the set of strings 
\begin{equation} \label{eqn:P}
  {\cal L} ~=~ \{~ x \in \{0,1\}^* ~:~ C_{|x|}(x) = 1 ~\}.
\end{equation}
Here $C_n(x)$ denotes the output of circuit $C_n$ 
when the gates are evaluated after input string $x$ (of length $n$) is used to load boolean values into the `input' vertices.
Then whenever the family is described by a logspace Turing machine, we have that ${\cal L} \in {\mathbf P}$.  Such families are called \emph{uniform}, and are always polynomially bounded \cite{book:Pap}. 

There are many ways to introduce randomness to this model.  For example, one may allow for extra vertices of zero indegree that are to be initialised randomly, or one may incorporate `coin-flip' vertices with indegree 1 that `overwrite' data with random data.  Within models which allow for some randomness, we redefine $C_{n}(x)$ to be the \emph{probability} of the output vertex evaluating to ``0''.   If there's a guarantee that $C_{|x|}(x) \not\in (\frac13,\frac23)$, for all strings $x$, then a language in {\bf BPP} may be derived.  This class is a semantic class since it depends on a guarantee.  There is nothing special about the $(\frac13,\frac23)$ limitation in that guarantee, since any non-negligible separation in the probabilities can be amplified by repeated computation \cite{book:Pap}.

Quantum circuits are usually defined similarly \cite{book:NandC} with unitary gates replacing ordinary vertices, and measurement gates relpacing output vertices.  Unitary gates have the same indegree as outdegree, because they are reversible.  Measurements are usually taken to be single-qubit measurements in the computational basis, (see \cite{book:NandC} for more details.)  Thus is the class {\bf BQP} derived.  
As well as being aptly notated pictorially by a directed acyclic graph, a quantum circuit can equally well be notated simply as a sequence of the unitary operators applied, which is to say we can describe it by listing a series of unitary elements from ${\cal A}_w$, to be applied sequentially to some starting input.  Without loss of generality, all measurement can be delayed to the end of the quantum circuit, and since only one bit of output is required for decision problems, that output measurement can then be a measurement of $Z$ on the first qubit, with other qubits being traced out.

\subsection*{DQC1 Circuits}
For DQC1 circuits we approach the problem slightly differently, because we need a different notion of `circuit input' to satisfy the DQC1 framework of allowable states.  Because the actual quantum input into the circuit must necessarily be of the form prescribed by equation (\ref{eqn:startstate}) for DQC1 circuits, we need an alternative way of getting the input string $x$ into the computation.  So we recycle the idea of a circuit being a list of unitary operators from ${\cal A}_w$, but  each element of the list will  be a \emph{choice of two} unitary operators, the actual one to apply being selected by a (classical) bit of $x$.
Thus an example of a description of a small DQC1 circuit would be $(A\backslash B)_{x_1} \cdot (C\backslash D)_{x_2} \cdot (E\backslash F)_{x_1}$.  This circuit, on input $x=~$``01'', would perform the automorphism specified by conjugation by the unitary $A \cdot D \cdot E$ on the starting state, but on input $x=$``11'' would conjugate the starting state by $B \cdot D \cdot F$, \emph{\&c}.
In general, we allow a polynomially bounded circuit width, $w$, so $A,B,C,D,E,F$ must be unitary elements of ${\cal A}_{w}$.
(Note that this definition of input could equally well have done for the classes described above, though it is not commonly employed.)

The circuit output probability $C_n(x)$ is then taken to be $(1+\beta)/2$ as per equations (\ref{eqn:genericDQC1state}) and (\ref{eqn:probability}), where $\beta$ is found from the state resulting from the starting state (equation (\ref{eqn:startstate})) being subject to the specified sequence of automorphisms.  As before, $C_n(x)$ has the physical meaning of being the probability of measuring the first qubit of the final state to be $\ket0$.
(An example of a problem that fits this paradigm -- indeed the canonical example -- is that of estimating the trace of a unitary matrix.  The input to the problem will amount to a classical description of the matrix.  The details are described below in section~\ref{sect:completeness}.)

Thus is the class {\bf BQ1P} derived\footnote{This is our notation, ``Bounded-error Quantum -- 1 pure qubit -- Polynomial time computation'', but we welcome correction from those who know better how to name new classes.}, assuming as before that the family of circuits is uniform.  
The precise rule that we adopt for defining {\bf BQ1P} languages interprets probabilities a little more generally too; instead of requiring that $C_{|x|}(x) \not\in (\frac13,\frac23)$ we simply require the (possibly) weaker condition that for $n=|x|$
\begin{equation} \label{eqn:polybounds}
  C_n(x) ~\not\in~ \left(~ \frac12-\frac{1}{2q(n)}, ~\frac12+\frac{1}{2q(n)} ~\right),
\end{equation}
for some known polynomially bounded complexity function $q(n)$.
The reason for allowing polynomially small error bounds is that there is no apparent way to `chain together' (sequentially compose) or even \emph{parallel compose} DQC1 circuits, and so unlike the {\bf BPP} or {\bf BQP} cases, amplification of probability has to take place \emph{outside} of the model.
And so any language in {\bf BQ1P} must satisfy  
\begin{eqnarray} \label{eqn:BQ1P}
  {\cal L}     &:=& \left\{~ x \in \{0,1\}^* ~:~ \beta(x) ~\ge~ \frac{1}{q(|x|)} ~\right\},  \\
  \bar{\cal L} &:=& \left\{~ x \in \{0,1\}^* ~:~ \beta(x) ~\le~ \frac{-1}{q(|x|)} ~\right\}. 
    \nonumber
\end{eqnarray}

The class is reasonably robust against variation in the exact specification of which unitaries are allowed within the circuit.  Since the circuit families must be uniform (\emph{i.e.} describable by a well-behaved logspace Turing machine,) it will be important to avoid encoding any `complexity' in the alphabet of allowable unitaries.  In general we will take a finite alphabet of gates and then work with its closure under conjugation by SWAP gates.  The resulting alphabet (subset of ${\cal A}_w$) will be of polynomially bounded size, and we shall want for it to form a universal set in the sense of being able to generate any unitary to arbitrary precision as $w \rightarrow \infty$.  An example of a such an alphabet would be the set of all CNot unitaries $~\frac{1+Z_j+X_k-Z_jX_k}{2}~$ together with the set of all Hadamards $~\frac{X_j+Z_j}{\sqrt2}~$ and all `$\pi/8$' unitaries $~\exp(iZ_j\pi/8)$.  Robustness of classes with respect to changes in the defining gate-set is discussed amply in the literature (\emph{e.g.}~\cite{book:NandC}.)

\section{Completeness} \label{sect:completeness}
Although {\bf BQ1P} is a semantic class, we can still provide a notion of completeness for it, by reference to the problem of estimating the (sign of the real part of the) trace of a unitary operator.  This problem is discussed in \cite{lit:KL98}, and we recap and expand on the main ideas here.

Formally, the promise problem for Trace Estimation is~: given a description of a unitary matrix in factored form (see below), decide whether the sign of the real part of the trace of the matrix is positive, given the promise that the magnitude of the real part of the trace is polynomially bounded away from zero.  Specifically, this means that if $x$ is the description of the unitary $U$, and $U$ belongs to the algebra ${\cal A}_w$, then the promise holds that $\bigl|~Re[Tr[U]]~\bigr| \ge \frac{2^w}{p(|x|)}$ for some polynomially bounded complexity function $p$.  The dependence on $w$ serves to make the problem well-defined, essentially remedying the fact that Trace applies a linear multiplier of $2^w$.  It may be noted \cite{lit:KL98} that for a randomly chosen $U$ without any such promise, the typical value of $\bigl|~Re[Tr[U]]~\bigr|$ is $\sqrt{2^w}$, and that in any case the value is constrained to the region $[0,2^w]$.  

Let $(L_{w,k})_{w,k}$ be some sensibly chosen \emph{uniform} family of maps that interpret binary strings of length $k$ into unitary elements of ${\cal A}_w$; (the details are unimportant.)
Let $~|x| = kt~$ and let 
\begin{equation} \label{eqn:decompU}
  U(x,w,k) ~=~ L_{w,k}(x_1 \ldots x_k) \cdots L_{w,k}(x_{kt-k+1} \ldots x_{kt}),
\end{equation}
denote the unitary matrix that is obtained by parsing the string $x$ in sections of $k$ bits, using $L_{w,k}$, and composing the resulting network of $t$ gates (each acting on $w$ qubits.)  

\begin{lemma}
  The promise problem for Trace Estimation is complete (using classical logspace reductions) for {\bf BQ1P}.
\end{lemma}
 
To show completeness of the promise version of the Trace Estimation problem for {\bf BQ1P} we need to do two things~: first we shall show that this version can be solved within the {\bf BQ1P} methodology, and second we shall show that any other methodology solving the promise problem can simulate the action of any other {\bf BQ1P} computation.  This will not however help us write down any particular language in {\bf BQ1P}, because the promise in question cannot be readily encoded in the usual linguistic syntax.  The notion of reduction that we use in reducing other problems to this Trace Estimation problem will be reduction by classical processing bound to logarithmic space \cite{book:Pap}; see lemma \ref{lem:reduction} in section \ref{sect:properties} for more details.

Proof part 1)~~(\emph{cf} \cite{lit:KL98}) For the first part of the proof, we need a uniform family of DQC1 circuits such that circuit number $(|x|,w,k)$, of width $w'$, on input $x,\rho_{\mbox{  \footnotesize{\!start}}}$, will generate a state for which $\beta(x) \ge 1/q(|x|)$ if and only if $Re[Tr[U(x,w,k)]] \ge 0$, \emph{etc}.
To achieve this, we take $w'=w+1$ and use a DQC1 circuit that begins and ends with a Hadamard gate on the first qubit, and in between these performs the relevant conditional unitaries on the remaining qubits, controlled also on the (quantum) setting of the first qubit to the state $\ket1$.
(The factors in equation~(\ref{eqn:decompU}) can be broken down in a uniform way\footnote{\emph{cf} section \ref{sect:properties}} into choices of unitaries controlled by the classical bits of $x$, and for any fixed alphabet, it is a finite problem to break down controlled versions of gates into simpler forms allowed in the alphabet, with arbitrary accuracy.)
Writing this out for any specific input $x$ gives
\begin{equation} \label{eqn:TraceEst}
  H_1 \cdot \Lambda_1(U(x,w,k)) \cdot H_1.
\end{equation}
This circuit, on input $\rho_{\mbox{  \footnotesize{\!start}}}$, will yield the ${\cal A}_{w+1}$ state 
\begin{equation}
  \frac{1 ~+~ \frac{U(x,w,k)+U^\dag(x,w,k)}{2} \cdot Z_1 ~-~ i \cdot \frac{U(x,w,k)-U^\dag(x,w,k)}{2} \cdot Y_1 }{2^{w+1}};
\end{equation}
for which the $\beta$ of equation (\ref{eqn:probability}) is $Re[~Tr[~ U(x,w,k) \in {\cal A}_w ~]/2^{w}~]$.
Invoking the promise, we can be sure that this $\beta$ satisfies $|\beta| \ge \frac{1}{p(|x|)}$, and so by taking $q>p$, we can ensure that the correct decision is made according to equation~(\ref{eqn:BQ1P}), even allowing room for small errors in any approximations that might be required when rendering equation~(\ref{eqn:TraceEst}).

Proof part 2)~~ For the second part of the completeness proof, suppose we have some `black box' means for deciding of the triple $(x,w,k)$ whether the real part of the trace of $U(x,w,k)$ is positive, given that it is bounded away from zero.  We shall use this black box, together with what amounts to merely logspace processing, to decide an arbitrary {\bf BQ1P} language.  The use of logspace processing will be `justified' later in section~\ref{sect:properties}.

An arbitrary {\bf BQ1P} language that we would have our black box decide may be written as 
\begin{equation} \label{eqn:genericDQC1Lang}
  \left\{ (y,w,k) 
          ~:~ Tr\left[ \frac{U(y,w,k) \cdot Z_1 \cdot U^\dag(y,w,k) \cdot Z_1}{2^{w}} \right] \ge \frac{1}{q(|y|)}  \right\},
\end{equation}
as derived from equations~(\ref{eqn:genericDQC1state}),~(\ref{eqn:probability}), and~(\ref{eqn:BQ1P}).  
Then all we need do is to have $x$ code for a related circuit of the same width and similar depth, so that $U(x,w,k) = U(y,w,k) \cdot Z_1 \cdot U^\dag(y,w,k) \cdot Z_1$ and $|x| = O(|y|)$; for then the black box will be able to decide on $(y,w,k)$ via $x$.  This will be a simple matter for any reasonably chosen $L_{w,k}$.  \QED

\section{Properties of DQC1 Circuits} \label{sect:properties}
This section gathers together some important properties of the circuits we've been discussing, and of {\bf BQ1P} itself.

\subsection*{Boolean formulas on inputs}
Observe that the following small DQC1 circuit will `compute' the boolean operation $x_1 \wedge x_2$, deterministically, using a circuit of total width $w \ge 1$~:
\begin{equation} \label{eqn:AND}
  (1\backslash H_1)_{x_1} \cdot (1\backslash Z_1)_{x_2} \cdot (1\backslash H_1)_{x_1}.
\end{equation}
More precisely, this will map the starting state to $(1+\beta Z_1)/2^w$ where $\beta = (-1)^{x_1 \wedge x_2}$.  Likewise, other simple boolean gates may be rendered.  
Note that it is necessary to use one of the bits twice (bit $x_1$ in this example) in order to perform a non-trivial gate.  In general, it would be exceptionally limiting if we were only allowed to use each input bit once, because after each input bit had been used, no further automorphisms would be able to change any of the algebraic relations held between the states arising from different $x$ values.

Does the ability to perform these simple boolean gates mean that it is straightforward to put {\bf P} $\subseteq$ {\bf BQ1P}?
Well, no, because there is no method prescribed to take the output of one DQC1 circuit directly into the input of another DQC1 circuit, and therefore it is not enough to be able to perform interesting gates only on \emph{input bits}, one must also be able to perform interesting gates on \emph{intermediate bits} (whatever they might be.)  Recall from equation (\ref{eqn:genericDQC1state}) that there is a somewhat limited `place' in which `intermediate data' can be held during a DQC1 computation.

\subsection*{Complement}
It is obvious from the definitions that {\bf co-BQ1P} = {\bf BQ1P}, and so we shall say no more about this, except to point out that the sign of $\beta$ in equation (\ref{eqn:genericDQC1state}) can be flipped at any time of our choosing by use of the unitary $X_1$.

\subsection*{$\bigoplus\mathbf{L} \subseteq \mathbf{BQ1P}$}
The well known complexity class $\bigoplus\mathbf{L}$, (see \emph{e.g.} \cite{lit:AG0406},) may be defined as comprising those languages that can be decided by a uniform family of circuits composed entirely of CNot gates, whereby the $|x|$th circuit decides on the string $x$ by acting on the pure state $\ket{x}$ prior to a (deterministic) measurement of the first qubit in the computational basis.
Equivalently, using our idea of ``input as classical control'', we could define $\bigoplus\mathbf{L}$ by having a uniform family of circuits composed entirely of classical-input-controlled CNot gates, in like fashion to the DQC1 circuit definition.  We can use this idea to show~:
\begin{lemma}  \label{lem:parityL}
  Class inclusion~:  $\bigoplus\mathbf{L} \subseteq \mathbf{BQ1P}$.
\end{lemma}

Suppose we had a uniform circuit family, the circuits being made up exclusively of CNot gates, (the bits of an input string $x$ being used as an additional control over some or all of those CNot gates,) and suppose those circuits were intended to act on initial \emph{pure} states of the form $\ket{100...00}$.  Suppose we were then to measure a given bit of the output.  We could describe the process by writing 
\begin{equation}
  C_{|x|}(x) ~~=~~ \bigl|~ \bra{0}_1 ~Tr_{[2..w]}[~ U_{|x|}(x)\ket{100...00} ~] ~\bigr|^2.
\end{equation}

Now suppose we were to make a dual family of circuits by taking every circuit from the original family and simply reversing the direction of each CNot gate therein, exchanging source qubit with target qubit, but retaining all of the classical controls on the gates from the bits in $x$.  Write the action of this modified circuit as $\widehat{U}_{|x|}(x)$.  We intend that circuits of this dual family $(\widehat{U}_n)_n$ be applied (with the same control from the input string $x$,) on the DQC1 state given in equation (\ref{eqn:startstate}).
It is easy to see that if an original circuit $U_{|x|}(x)$ would have mapped its pure starting state $\ket{100...00}$ to  $\ket{s_1s_2...s_w}$, say, where $s_1=1-C_{|x|}(x)$, then the dual circuit $\widehat{U}_{|x|}(x)$ would map its mixed starting state $\rho_{\mbox{  \footnotesize{\!start}}}$ to the state $(1+Z_1^{s_1}Z_2^{s_2}...Z_w^{s_w})/2^w$.  
This is because whereas $\Lambda_i(X_j) \cdot \ket{a}_i \ket{b}_j = \ket{a}_i \ket{a+b}_j$, we have things the other way around in DQC1 notation~: 
\begin{equation}
  \Lambda_i(X_j) \cdot Z_i^a \cdot Z_j^b \cdot \Lambda_i(X_j) ~=~ Z_i^{a+b} \cdot Z_j^b.
\end{equation}

Consider next the somewhat larger DQC1 circuit composed as $\widehat{U}_n^\dagger \cdot X_1 \cdot \widehat{U}_n$.  For a given control string $x$, this will map the state $\rho_{\mbox{  \footnotesize{\!start}}}$ to $\frac{1+\beta Z_1}{2^w}$, where $\beta = 2C_{|x|}(x) - 1$, as required, as the reader may readily check.  Thus for such circuit families, we can guarantee determinism in the final measurement, and so certainly a {\bf BQ1P} language is issued. 
This shows that $\bigoplus\mathbf{L} \subseteq \mathbf{BQ1P}$.   \QED

\subsection*{More pure qubits}
The next thing to show is that the complexity class {\bf BQ1P} remains the same if we allow a constant number of pure qubits instead on just one, or even a logarithmic number.

To be precise, suppose we generalise the model by having $c$ pure qubits and being allowed, after arbitrary unitary processing, to measure just whether the first $d$ qubits are \emph{all} in the state $\ket0$.  For each circuit $U(x)$ we define $\beta_{c,d}$ accordingly, (\emph{cf.} eqn~(\ref{eqn:probability})) ~:  
\begin{equation} \label{eqn:DQCcComp}
  \frac{1+\beta_{c,d}}{2} ~:=~ Tr\left[~ U(x) \cdot \frac{(1+Z)^{\otimes c}}{2^w} \cdot U(x)^\dag \cdot \frac{(1+Z)^{\otimes d}}{2^d} ~\right].
\end{equation}

There would then be no point in having $d-c > O( \log |x| )$, since the signal strength would drop off superpolynomially for all $U(x)$, and so there would be no way to satisfy the requirement of equation~(\ref{eqn:polybounds}) anywhere.  
If we were to have $c \sim d \sim O( \log n )$ then equation (\ref{eqn:DQCcComp}) can be expanded to be the scaled sum of the traces of polynomially many unitary operators, and thus, we shall show, the action of the machine could be simulated in the {\bf BQ1P} model.  
There are essentially two ways to complete the simulation in this case, either by parallel application of polynomially many simulators followed by some classical accounting at the end, or by randomly choosing one of those unitary operators to run the simulation on, to obtain a $\beta$ corresponding to the \emph{average} trace.  The first technique is not apparently applicable to {\bf BQ1P} methodology, because as we have already noted, there is no real notion of parallel composition in DQC1.  But the second option can be simulated in {\bf BQ1P}, and though it may provide a polynomially weaker signal, it won't be so weak as to violate the guarantee required by equation (\ref{eqn:polybounds}).  

Let us illustrate this simulation in more detail for the important case of $c=2, d=1,$ (using the temporary notation ``{\bf BQ2P}'' to indicate the class generated from allowing for two pure qubits instead of one) ~:
\begin{lemma}  \label{lem:classEquality}
  Class equality~:   {\bf BQ2P = BQ1P}
\end{lemma}

The probability that we wish to sample is given by 
\begin{eqnarray} \label{eqn:DQCcComp1}
  \frac{1+\beta_{2,1}(x)}{2} &=& Tr\left[~ U(x) \cdot \frac{(1+Z_1)(1+Z_2)}{2^w} \cdot U(x)^\dag \cdot \frac{(1+Z_1)}{2^1} ~\right] \nonumber \\
                           &=& \frac12 + \frac{\frac{Tr[U Z_1 U^\dag Z_1]}{2^w}+\frac{Tr[U Z_2 U^\dag Z_1]}{2^w}+\frac{Tr[U Z_1 Z_2 U^\dag Z_1]}{2^w}}2,
                             \nonumber \\  
\end{eqnarray}
where $U$ (and hence $\beta_{2,1}$) depends on the input string, $x$.
Now there is a DQC1 circuit of width $w+2s$ that gets very close to sampling the probability $\frac{1+\beta_{2,1}/3}{2}$, namely 
\begin{equation}  \label{eqn:MarkovCircuit}
  U(x) \cdot \Lambda_{1,w+2s}(X_2) \cdot \Lambda_{2,w+2s-1}(X_1) \cdots 
             \Lambda_{1,w+2}(X_2) \cdot \Lambda_{2,w+1}(X_1).
\end{equation}
This approximation works `exponentially well' because the Toffoli gates that source the extra $2s$ qubits are sourcing totally random qubits, and so they reduce to random CNot gates, which establish a rapid-mixing Markov process for simulating the alternative starting signal; this being an almost-uniform mix from the set $\{Z_1,Z_2,Z_1Z_2\}$ instead of the usual $Z_1$.  
So if there were a `\textbf{BQ2P} language' according to the semantic guarantee
\begin{equation}
  \bigl| \beta_{2,1}(x) \bigr| ~\ge~  \frac{1}{q(|x|)},
\end{equation}
then the same language would be in \textbf{BQ1P} according to the semantic guarantee
\begin{equation}
  \bigl| \beta(x) \bigr|       ~\ge~  \frac{1}{3q(|x|)} - \frac{1}{3\cdot4^{s-1}},
\end{equation}
as the reader may easily check by considering the effect of the random Toffoli gates of the circuit in equation~(\ref{eqn:MarkovCircuit}) on the starting state.
Recall that $q$ is polynomially bounded, so $s$ need only be logarithmic.  \QED

This illustration will readily generalise to show that class {\bf BQ1P} is not extended by increasing the number of pure input bits to a logarithmic amount, (nor indeed by increasing the number of output bits by any amount, under the measurement assumptions we made.)

\subsection*{Reduction}
We are now in a position to show that~:
\begin{lemma}  \label{lem:reduction}
  Class reduction~:  {\bf BQ1P} is closed under reduction by logspace deterministic processes.
\end{lemma}

This is the usual notion of reduction employed in complexity theory, \cite{book:Pap}.  It will in turn justify our earlier assumptions in section~\ref{sect:completeness} relating to reduction concepts.

The idea is to use lemmas~\ref{lem:parityL} and~\ref{lem:classEquality}, so that if ${\cal L} \in $ {\bf BQ1P} and $R$ is a function whose every bit is in $\bigoplus\mathbf{L}$, (which is a more general thing than merely saying that $R$ is logspace deterministic computable,) then we can prove the `pre-reduction' language ${\cal L}' = \{~ x : R(x) \in {\cal L} ~\}$ to be in `\textbf{BQ2P}', and hence in \textbf{BQ1P}.  

So proceed by supposing that we do have the use of two pure qubits rather than just one, and then divide the available qubits into two partitions with one pure qubit in each partition.  The first partition will be used to undergo the transformations of the circuit being implemented, while the second partition will be used to compute the bits of $R$ as they are needed.  Each time bit $R_i(x)$ is needed to choose between unitaries for applying to the first partition, use the techniques described for lemma~\ref{lem:parityL} to map the second partition from state $\rho_{\mbox{  \footnotesize{\!start}}}=(1+Z_1)/2^w$ to state $(1+(-1)^{R_i(x)}Z_1)/2^w$.  Then apply the unitary on the first partition \emph{controlled on} the pure qubit in the second partition, before `uncomputing' the computed control bit to restore the second partition to its original $\rho_{\mbox{  \footnotesize{\!start}}}$ value.  (The partitions do not become entangled nor even classically correlated, because the control is always on a bit known to be pure and in the computational basis.)
Apply lemma~\ref{lem:classEquality} to finish the proof.  \QED
\medskip

The closure under logspace reduction, stability under small pertubations to the model, and the various completeness properties, give us some reassurance that the class has been defined correctly in a `natural' way.

\section{Computational complexity conjectures} \label{sect:conjectures}
We subscribe to the school of thought which assumes that each of {\bf BPP} $\not=$ {\bf BQP} and {\bf L} $\not=$ {\bf P} is true but too difficult to prove.  Consistent with this belief, here are some other conjectures~:

\begin{conjecture} \label{conj:1}
  {\bf P} $\not\subseteq$ {\bf BQ1P}
\end{conjecture}

\begin{conjecture} \label{conj:2}
  {\bf BQ1P} $\not\subseteq$ {\bf BPP}
\end{conjecture}

\begin{conjecture} \label{conj:3}
  $\mathbf{BQP} \not\subseteq \mathbf{BPP}^{\mathbf{BQ1P}}$~: 
  No hybrid machine consisting of classical processors and quantum-uncorrelated DQC1 processors can convincingly simulate the action of a quantum Turing machine within polynomial time, even allowing for use of adaptive techniques (classical correlations.)
\end{conjecture}

Clearly these conjectures are too strong to hope to prove affirmatively, given the belief expressed above.  
For example, a proof of conjecture~\ref{conj:1} would immediately establish that $\bigoplus\mathbf{L}$ (and hence also {\bf L}) is a proper subset of {\bf P}, while a proof of conjecture~\ref{conj:2} or~\ref{conj:3} would confirm our belief that quantum computation is indeed superpolynomially more powerful than its classical counterpart.  Conjecture~\ref{conj:1} is important, since a `straw poll'\footnote{Ok, my methods aren't always especially scientific, and I'm not admitting in print just how many experts confessed a strong opinion on this matter.} reveals that opinion is divided as to whether it is even \emph{likely}, let alone provable.  A popular approach is to conjecture that {\bf BQ1P} = {\bf BQP}, and so one should make a serious attempt to look for evidence to resolve this discrepancy.  Evidence for conjecture~\ref{conj:2} can be found in section \ref{sect:completeness}, since the problem of Trace Estimation appears to be classically hard.  Conjecture~\ref{conj:3} is a bit more speculative and intuitive, capturing our opinion that having access to many pure qubits in the \emph{same} processor really is a valuable resource.
\medskip

\subsection*{Evidence by relativisation}
Our main reason for finding conjectures~\ref{conj:1} and~\ref{conj:2} plausible has to do with relativisation, using oracles.  
The standard model for oracles fits into the DQC1 model only by representing a classical oracle as an (unknown) non-uniform class of permutation matrices which we may use as building blocks within the circuits in a circuit family, but which we may not modify nor examine any description of.\footnote{We assume that we are allowed to implement certain derivatives within the circuit, such as ``controlled-$U$'', a polynomial number of times if necessary.}  
The idea is that the complexity of computation represented by that oracle can be utilised `atomically' within the circuitry, despite the fact that DQC1 doesn't allow for direct concatenation of computations, nor `inputs and outputs' in the traditional sense, (as discussed in the first section.)

Let $U$ refer to such a ($2^w$-by-$2^w$) matrix.  Then it is clear that while {\bf P} circuitry relativised to $U$ can report back any entry of the matrix, the best that {\bf BQ1P} circuitry can do (see section \ref{sect:completeness}) is to estimate $\beta = Tr[V_U]/2^{w'}$, where $V_U$ is some polynomial-length unitary `word' in the algebra ${\cal A}_{w'}$, some of whose `letters' may be the unknown $U$.  
That there is a definite separation here can be shown inductively.  Let $U$ be a unitary matrix with a $(\phantom|_0^1\phantom|_1^0\phantom|)$ in the top left corner and let $U'$ be the same matrix but with the top left corner replaced by $(\phantom|^0_1\phantom|^1_0\phantom|)$, \emph{i.e.} $U' = U - 2\ketbra--$.  Whereas {\bf P} circuitry is able to distinguish between these two oracles, we shall show that the difference $~Tr[V_U] - Tr[V_{U'}]~$ cannot be significant, \emph{i.e.} cannot be on the order $2^{w'}/poly(w)$.
First consider the case $w'=w$ and use induction on the `word-length index' $t$ found in the expression $V_{U,t} = A_t \cdot U \cdot A_{t-1} \cdot U \cdots U \cdot A_1$.
Now define
\begin{equation}
  \chi_t ~~:=~~ \max_W ~\bigl|~ Tr[~ W \cdot ( V_{U,t} - V_{U',t} ) ~] ~\bigr|, \quad \mbox{$W$ unitary.}
\end{equation}
From here we see
\begin{eqnarray}
  \lefteqn{ \bigl|~ Tr[~V_{U,t}~] ~-~ Tr[~V_{U',t}~] ~\bigr| } && \\
      &\le& \chi_t  \nonumber \\
      &=&   \bigl|~ Tr[~ W \cdot A_t \cdot ( U \cdot V_{U,t-1} - U' \cdot V_{U',t-1} ) ~] ~\bigr|  \nonumber \\
      &=&   \bigl|~ Tr[~ ( W A_t U ) ( V_{U,t-1} - V_{U',t-1} ) ~] ~+~ 2\bra- V_{U',t-1} W A_t \ket- ~\bigr|  \nonumber \\
      &\le& \chi_{t-1} + 2, \nonumber
\end{eqnarray}
and so this difference between the traces cannot grow to the exponential required within polynomial word-length $t$.  The argument is very similar for $w'>w$, so we shall leave details to the reader.

Therefore, if ${\cal O}$ is an oracle containing infinitely many (randomly chosen) examples of $(U,U')$ pairs, then (almost surely) one can construct deterministic distinguishing algorithms in {\bf P} that have no analogue in {\bf BQ1P}, and hence an oracle-based decision language in the former that cannot be contained in the latter, so $\mathbf{P}^{\cal O} \not\subseteq \mathbf{BQ1P}^{\cal O}$.

\medskip
For conjecture~\ref{conj:2}, consider taking $U$ to be a random $2^w$-by-$2^w$ permutation matrix, implementing the permutation $\pi$ on $GF(2)^w$, subject to a certain promise (to be specified shortly,) and let $V = H^{\otimes w} \cdot U \cdot H^{\otimes w} \cdot U \cdot H^{\otimes w} \cdot U$.  
Then
\begin{eqnarray}
  \frac{Tr[V]}{2^w}  
    &=& 2^{-5w/2} \sum_{ikm} (-1)^{i \cdot \pi(k) ~\oplus~ k \cdot \pi(m) ~\oplus~ m \cdot \pi(i)}  \\
    &=& 2^{w/2}\bigl(~ 1 - 2~\mathbbm{E}[~i \cdot \pi(k) ~\oplus~ k \cdot \pi(m) ~\oplus~ m \cdot \pi(i)~] ~\bigr),
  \nonumber 
\end{eqnarray}
and provided we have the promise that this value is sufficiently bounded away from 0, it is easy for {\bf BQ1P} to resolve whether the value is closer to 1 or $-1$, but clearly any (randomised) classical device would have to query $U$ at exponentially many places to establish that distinction.  
Therefore any ${\cal O}$ containing infinitely many (randomly chosen) examples of such $U$ matrices will (almost surely) exemplify a case where $\mathbf{BQ1P}^{\cal O} \not\subseteq \mathbf{BPP}^{\cal O}$.

\section*{Acknowledgements}
Our thanks go to Richard Jozsa and Aram Harrow, for much proof-reading and encouragement.  This work has been sponsored by CESG, (UK Government.)

\end{document}